\documentclass[%
reprint,
superscriptaddress,
bibnotes,
amsmath,amssymb,
prb,
floatfix,
]{revtex4-2}

\usepackage{graphicx}% Include figure files
\usepackage{dcolumn}% Align table columns on decimal point
\usepackage{bm}% bold math
\usepackage{hyperref}% add hypertext capabilities
\usepackage{xcolor}
\usepackage{tikz}

\definecolor{deggreen}{rgb}{0.544,0.8,0.16}
\definecolor{degblue}{rgb}{0.16,0.544,0.8}

\DeclareRobustCommand{\excdeg}{%
    \begin{tikzpicture}%
        \filldraw[fill=deggreen,draw=black] circle (2.4pt);
    \end{tikzpicture}%
}

\DeclareRobustCommand{\biexcdeg}{%
    \begin{tikzpicture}%
        \filldraw[fill=degblue,draw=black,rotate=45] rectangle (4pt,4pt);
    \end{tikzpicture}%
}

\newcommand{\ket}[1]{\left| #1 \right\rangle}

\begin{document}

\preprint{}

\title{Biexcitons in Highly Excited CdSe Nanoplatelets}% Force line breaks with \\
%\thanks{A footnote to the article title}%

\author{F. Garc\'ia Fl\'orez}
\email{f.garciaflorez@uu.nl}
\affiliation{Institute for Theoretical Physics and Center for Extreme Matter and Emergent Phenomena, Utrecht University, Princentonplein 5, 3584 CC Utrecht, The Netherlands}

\author{Laurens D. A. Siebbeles}
\email{l.d.a.siebbeles@tudelft.nl}
\affiliation{Optoelectronic Materials Section, Department of Chemical Engineering, Delft University of Technology, Van der Maasweg 9, 2629 HZ, Delft}

\author{H. T. C. Stoof}
\email{h.t.c.stoof@uu.nl}
\affiliation{Institute for Theoretical Physics and Center for Extreme Matter and Emergent Phenomena, Utrecht University, Princentonplein 5, 3584 CC Utrecht, The Netherlands}

\date{\today}% It is always \today, today,
%  but any date may be explicitly specified

\begin{abstract}
    We present the phase diagram of free charges (electrons and holes), excitons, and biexcitons in highly excited CdSe nanoplatelets that predicts a crossover to a biexciton-dominated region at easily attainable low temperatures or high photoexcitation densities.
    Our findings extend previous work describing only free charges and excitons by introducing biexcitons into the equation of state, while keeping the exciton and biexciton binding energies constant in view of the relatively low density of free charges in this material.
    Our predictions are experimentally testable in the near future and offer the prospect of creating a quantum degenerate, and possibly even superfluid, biexciton gas.
   Furthermore, we also provide simple expressions giving analytical insight into the regimes of photoexcitation densities and temperatures in which excitons and biexcitons dominate the response of the nanoplatelets.
\end{abstract}

\maketitle

%\tableofcontents
%\newpage

\section{\label{sec:introduction}Introduction}

Biexcitons, that is, bound states of two excitons, have been extensively studied in the literature both theoretically and experimentally in various materials and under different conditions \cite{lampert1958,ungier1989,banyai1987,wang2020,you2015,takayama2002,seller2019,pei2017,thouin2018,steinhoff2018,ronnow2012,bacher1999,kunneman2014,guzelturk2014,bouet2014}.
In particular, the study of the optical properties of excitons and biexcitons has received much attention in recent years, first in order to understand the underlying physics, and later towards possible new applications \cite{takagahara1989,phillips1992,nair2011,marz1980,ivanov1993,neukirch2000,geiregat2019,zhang2014,britnell2013,bernardi2013,lopez-sanchez2013,liu2015}.
Of special interest are biexciton-mediated lasing applications, due to the benefits brought about by the characteristics of biexcitons such as low thresholds and room-temperature utilization as a result of enhanced Coulomb interactions in low dimensions \cite{booker2018,masumoto1993,grim2014,soavi2016,wang2019,shang2017,salehzadeh2015,wu2015,yang2017,guzelturk2019,li2017,pelton2018}.
Furthermore, ever since two-dimensional materials became easily manufactured, for instance by chemical vapor deposition and colloidal self-assembled growth, they have been increasingly the object of study for the development of optoelectronic devices \cite{schwarz2014,mak2010,murray1993,chhowalla2013,singh2018,ross2014,yin2005,mahler2010,mak2010,zhao2018,mahler2012,ithurria2011}.

\begin{figure}%[h!]
    \begin{center}
        \includegraphics[width=\linewidth]{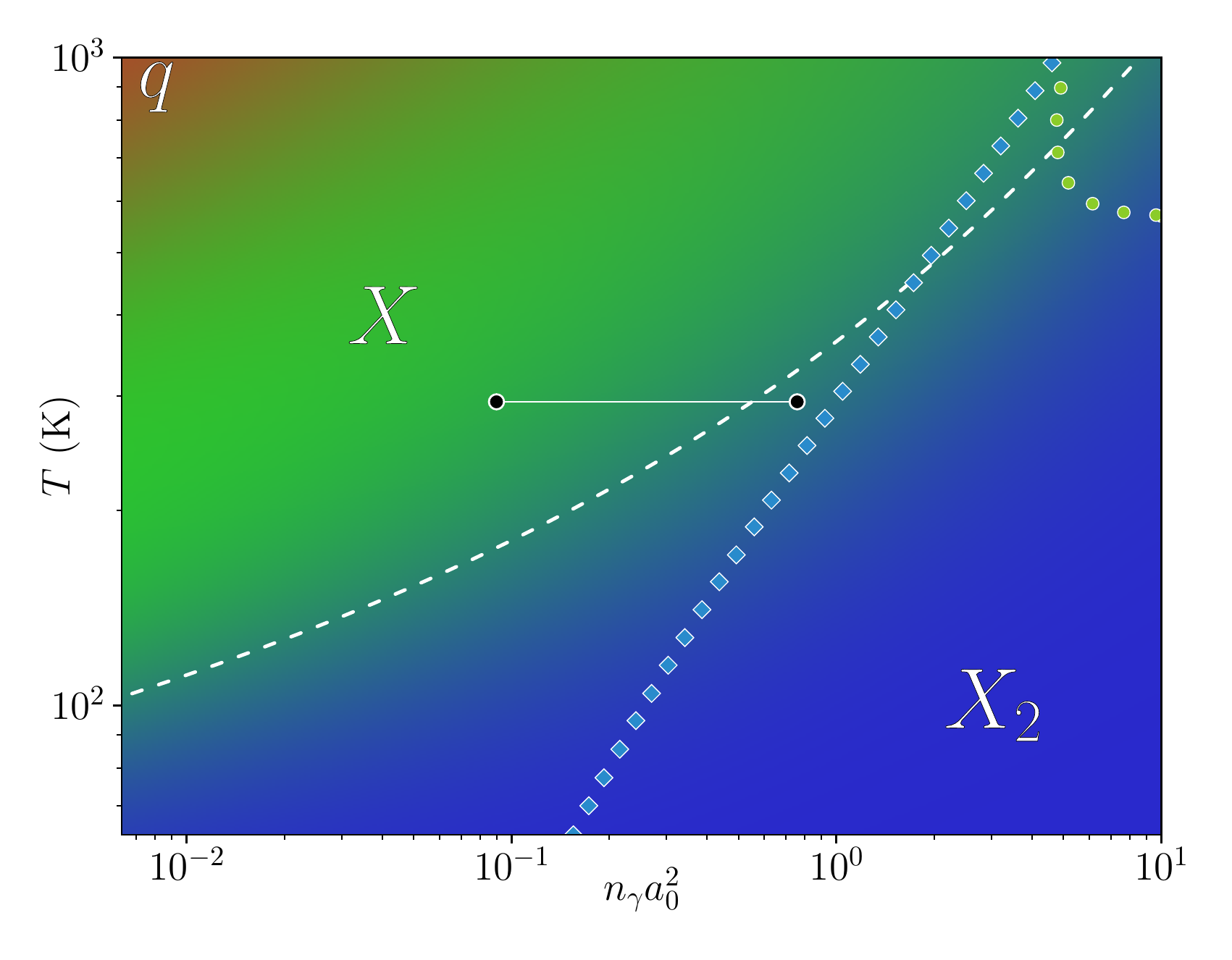}
        \caption{
            Fraction of free charges (red, top left, label $q$), excitons (green, middle, label $X$), and biexcitons (blue, bottom right, label $X_2$), as a function of temperature $T$ and photoexcitation density scaled by the (three-dimensional) exciton Bohr radius squared $n_\gamma a_0^2$.
            The white horizontal solid line corresponds to the region experimentally explored in Ref.\ \cite{tomar2019}, at the constant temperature $T=294$\ K.
            The white dashed line marks the points for which the density of excitons and biexcitons is equal.
            The circles (\excdeg, top right) and squares (\biexcdeg, from bottom to top right) dotted lines represent the photoexcitation density above which excitons and biexcitons become quantum degenerate, respectively.
        }
        \label{fig:diagram}
    \end{center}
\end{figure}

Even though excitons and biexcitons have been thoroughly experimented with, it was not until recently that an unexpected stability of excitons in CdSe nanoplatelets at high densities was observed, which was in turn explained by the rather unimportant screening effects of the free charges \cite{tomar2019,garciaflorez2019}.
Armed with this newfound understanding, it follows that the formation of more complex species is very likely as well.
Having established that screening alone does not unbind excitons, which would result in an electron-hole plasma regime, the thermodynamical description of the phase diagram of highly excited CdSe nanoplatelets may be extended via the introduction of another species: an ideal gas of biexcitons, described similarly as excitons.
Biexcitons are complexes that form due to the not particularly strong attraction between excitons, represented by the symbol $X$, which are analogous to hydrogen molecules and in turn are represented by $X_2$.
In order to obtain results suitable for experimental exploration, we restrict our discussion to pump-probe experiments, in which the pump laser optically excites electrons and holes equally, that is, $n_q \equiv n_e = n_h$ where $n_q$ is half the density of free charges, and $n_e$ and $n_h$ are the density of electrons and holes, respectively.
After a short period of thermalization the system reaches a chemical (quasi) equilibrium regime, in which a given photoexcitation density thus satisfies $n_\gamma \equiv n_q + n_X + 2 n_{X_2}$, where $n_X$ is the density of excitons, and $n_{X_2}$ is that of biexcitons.

With the purpose of better understanding the overall picture regarding the fraction of free charges $q$, excitons $X$, and biexcitons $X_2$ we present in Fig. \ref{fig:diagram} a phase diagram in terms of temperature $T$ and photoexcitation density scaled by the exciton Bohr area $n_\gamma a_0^2$, color coded to represent each species.
Figure \ref{fig:diagram} shows three different regions in which each species dominates, following a clear trend from high temperatures and low density to low temperatures and high density, with a smooth crossover connecting each region.
It is promising for experiments that merely lowering the temperature from $T \simeq 300$\ K to just $T \simeq 100$\ K at the densities explored by Ref.\ \cite{tomar2019} is well enough to reach the biexciton-dominated regime.
Moreover, the circles (\excdeg) and squares (\biexcdeg) dotted lines represent the photoexcitation density and temperature at which excitons and biexcitons become quantum degenerate, that is, they represent (a lower bound for) the Kosterlitz-Thouless transition to the superfluid regime at high enough densities, thus requiring the use of quantum statistics for the bosonic species of interest \cite{kosterlitz1973}.

Our paper is organized as follows.
Section \ref{sec:eq_state} presents a thermodynamical description of the system, which we use then in Sec. \ref{sec:phase} to compute the density of free charges $n_q$, density of excitons $n_X$, and density of biexcitons $n_{X_2}$ as a function of the photoexcitation density $n_\gamma$ and temperature $T$.
Lastly, Sec. \ref{sec:discussion} summarizes our findings, delves into some of the approximations used, and gives an outlook on future endeavors.

\section{\label{sec:eq_state}Equation of state}

Given that excitons stay bound  even at rather high densities, a simple thermodynamical model involving three coupled ideal gases (electrons, holes, and excitons) is an adequate description for the photoexcitation densities at room-temperature explored by Ref.\ \cite{tomar2019}.
However, at higher photoexcitation density or lower temperature, a fourth species has to be incorporated: the biexcitons.
Because experimental results show that screening effects are negligible in CdSe nanoplatelets, we provide a consistent approach by considering both the energy level of the exciton state $E_X$ and that of the biexciton state $E_{X_2}$ as constants, i.e., independent of the density of free charges.
This is justified by the result for excitons obtained in Ref.\ \cite{garciaflorez2019}, where it was found that at room temperature $E_B$ varies only from -193 meV to -177 meV for the photoexcitation densities considered.
Note that these energies are by definition negative.
By coupling together well-known expressions for the density of an ideal gas given its chemical potential, we define the following thermodynamical equilibrium model.
It includes two ideal gases of fermions, i.e., electrons and holes, and two ideal gases of bosons, i.e., excitons and biexcitons.
Writing the equations for the density of each species explicitly, we have

\begin{align}
    \begin{split}
        \label{eq:id_gas_n_q}
        n_q &= g_s \frac{m_e k_B T}{2\pi \hbar^2} \ln\left(1 + e^{\mu_{\text{e}} / k_B T}\right) \\
            &= g_s \frac{m_h k_B T}{2\pi \hbar^2} \ln\left(1 + e^{\mu_{\text{h}} / k_B T}\right) ~~,
    \end{split} \\
    \begin{split}
        \label{eq:id_gas_n_exc}
        n_{X} &= -g_s^2 \frac{m_X k_B T}{2\pi \hbar^2} \\
              &\phantom{=}~ \times \ln\left(1 - e^{\left(\mu_{X} - E_{X}\right) / k_B T}\right) ~~,
    \end{split} \\
    \begin{split}
        \label{eq:id_gas_n_biexc}
        n_{X_2} &= - \frac{m_{X_2} k_B T}{2\pi\hbar^2} \\
                &\phantom{=}~ \times \ln\left( 1 - e^{\left(2\mu_{X} - 2 E_X - E_{X_2}\right)/k_B T} \right)
    \end{split}
    ~~.
\end{align}

\noindent Here $m_X = m_e + m_h$ is the mass of the exciton and $m_{X_2} = 2 m_X$ that of the biexciton.
The chemical potential for excitons is $\mu_X = \mu_{\text{e}} + \mu_{\text{h}}$, and that of biexcitons is $\mu_{X_2} = 2 \mu_X$.
Furthermore, these densities together give the photoexcitation density as $n_\gamma \equiv n_q + n_X + 2 n_{X_2}$.
Here we have introduced the number of degenerate spin states, that is, with the same energy, $g_s = 2$ of electrons ($\ket{\uparrow}$ and $\ket{\downarrow}$) and holes ($\ket{\Uparrow}$ and $\ket{\Downarrow}$), and correspondingly $g_s^2 = 4$ for excitons ($\ket{\uparrow\Uparrow}$, $\ket{\uparrow\Downarrow}$, $\ket{\downarrow\Uparrow}$, and $\ket{\downarrow\Downarrow}$).
Notice that in the case of biexcitons we consider instead a non-degenerate ground state, analogously to hydrogen molecules for which only one combination of spins corresponds to the ground state, that is, the singlet-singlet combination $(\ket{\uparrow\downarrow}-\ket{\downarrow\uparrow})\times(\ket{\Uparrow\Downarrow} - \ket{\Downarrow\Uparrow}) / 2$.
In order to correctly account for the quantum behavior of excitons and biexcitons, Eqs. (\ref{eq:id_gas_n_q}), (\ref{eq:id_gas_n_exc}), and (\ref{eq:id_gas_n_biexc}) are obtained using quantum statistics, that is, using the Fermi-Dirac and Bose-Einstein distributions.
Section \ref{sec:phase} presents a more in-depth discussion showing that at high enough photoexcitation density both excitons and biexcitons are correctly described only by using quantum statistics.

In order to stay close to experiments, our results are obtained for the CdSe nanoplatelets of 4.5 monolayers studied by Ref.\ \cite{tomar2019}, which sets $m_e = 0.27 ~ m_0$ and $m_h = 0.45 ~ m_0$ as the effective electron and hole masses, with $m_0$ the bare electron mass, corresponding to the $n=4$ case in Ref.\ \cite{benchamekh2014}.
Concerning the exciton and biexciton energy levels, these are fixed to the low density values measured by Ref.\ \cite{tomar2019}, i.e., $E_X = -193$\ meV and $E_{X_2} = -45$\ meV, with the corresponding (three-dimensional) exciton Bohr radius $a_0 = 2.16$\ nm computed as in Refs. \cite{garciaflorez2019,florez2020}.
Note again that we do not consider changes in $E_X$ due to screening by free charges, as shown by the experimental results obtained for CdSe nanoplatelets.
Since excitons do not break down even at rather high photoexcitation density, screening does not change the qualitative picture presented in Fig. \ref{fig:diagram}, and consequently our prediction is not significantly affected by expectedly lower values of $E_X$ or $E_{X_2}$.

Because of the equal density of electrons and holes, Eq. (\ref{eq:id_gas_n_q}) yields for the relation between the chemical potential of electrons $\mu_{\text{e}}$ and that of the holes $\mu_{\text{h}}$ that

\begin{equation} \label{eq:chem_pot_rel}
    \mu_{\text{h}}(\mu_{\text{e}}) = k_B T \ln\left(\left(1 + e^{\mu_{\text{e}} / k_B T}\right)^{\frac{m_e}{m_h}} - 1\right)
    ~~,
\end{equation}

\noindent where we have used

\begin{equation} \label{eq:chem_pot_density}
    \mu_{\alpha}(n_q) = k_B T \ln\left( e^{\frac{\pi\hbar^2}{m_\alpha k_B T}n_q} - 1 \right)
\end{equation}

\noindent for $\alpha =$ e, h.
Notice that in the case of $m_e = m_h$, both chemical potentials reduce to the same value.
Moreover, because each photon creates exactly one electron-hole pair the photoexcitation density $n_\gamma$ is half the total density of charges, both free and bound in excitons and biexcitons.
Therefore, a given value of the photoexcitation density at a fixed temperature corresponds to a specific value for each density $n_q$, $n_X$, and $n_{X_2}$.

\section{\label{sec:phase}Phase diagram}

Let us analyze in more detail the phase diagram presented in Fig. \ref{fig:diagram}.
Each species is represented with a color: red for free charges, green for excitons, and blue for biexcitons, with labels $q$, $X$, and $X_2$ respectively.
Then, the final color-coded plot is obtained from the fraction of each species $\mathcal{Q}$ in relation to the others, labeled by $\mathcal{X}_{\mathcal{Q}}$ and defined as

\begin{equation}
    \mathcal{X}_{\mathcal{Q}} \equiv \frac{n_{\mathcal{Q}}}{n_q + n_X + n_{X_2}}
    ~~,
\end{equation}

\noindent where $\mathcal{Q}$ is $q$, $X$, or $X_2$.

\begin{figure}[h!]
    \begin{center}
        \includegraphics[width=\linewidth]{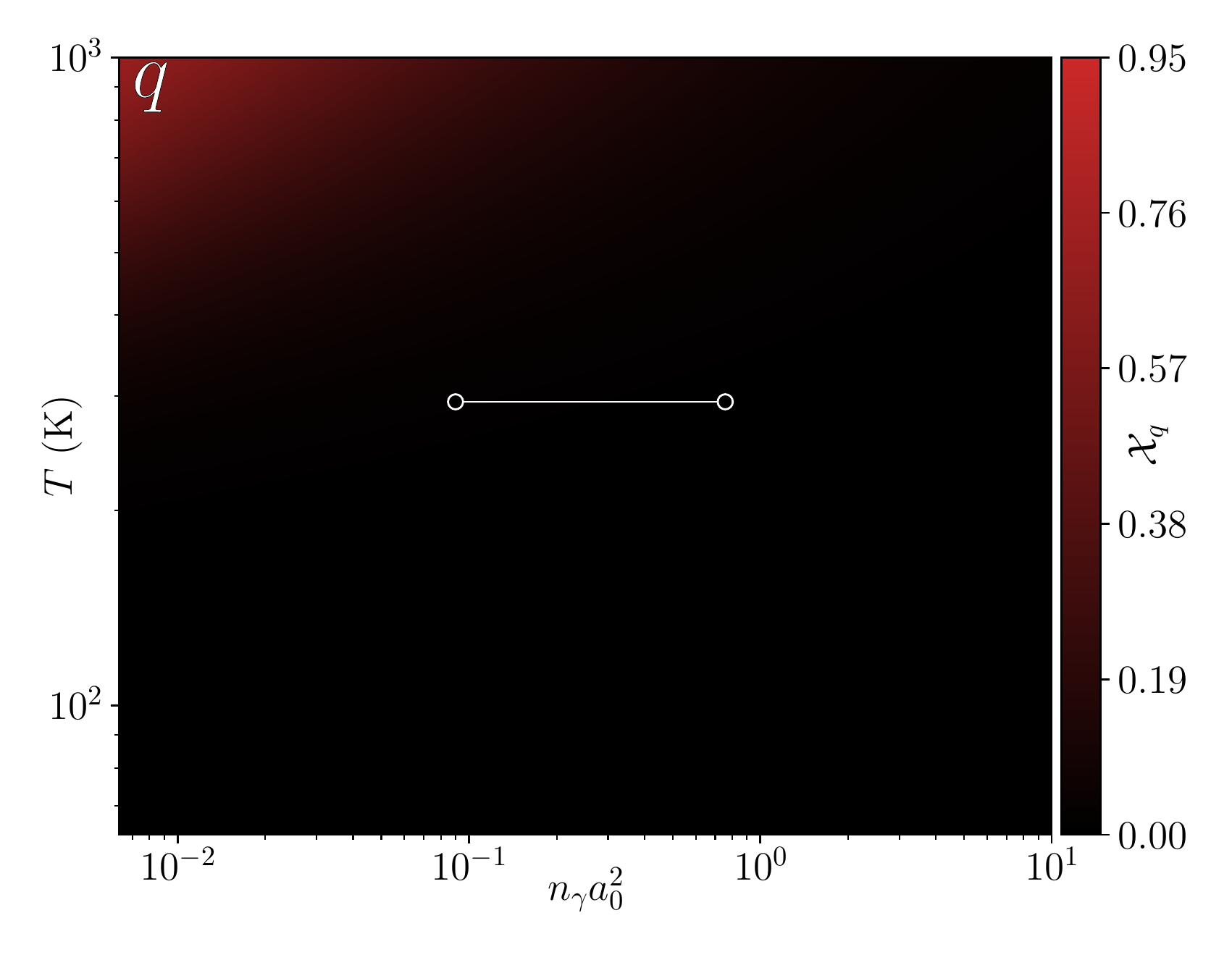}
        \caption{
            Fraction of free charges $\mathcal{X}_q$ as a function of temperature $T$ and photoexciation density scaled by the exciton Bohr area $n_\gamma a_0^2$.
            The maximum number of free charges per Bohr area is $n_q a_0^2 \simeq 0.145$, with $a_0^2 = 4.68$\ nm$^2$, obtained in the upper right corner.
        }
        \label{fig:diagram_n_q}
    \end{center}
\end{figure}

\begin{figure}[h!]
    \begin{center}
        \includegraphics[width=\linewidth]{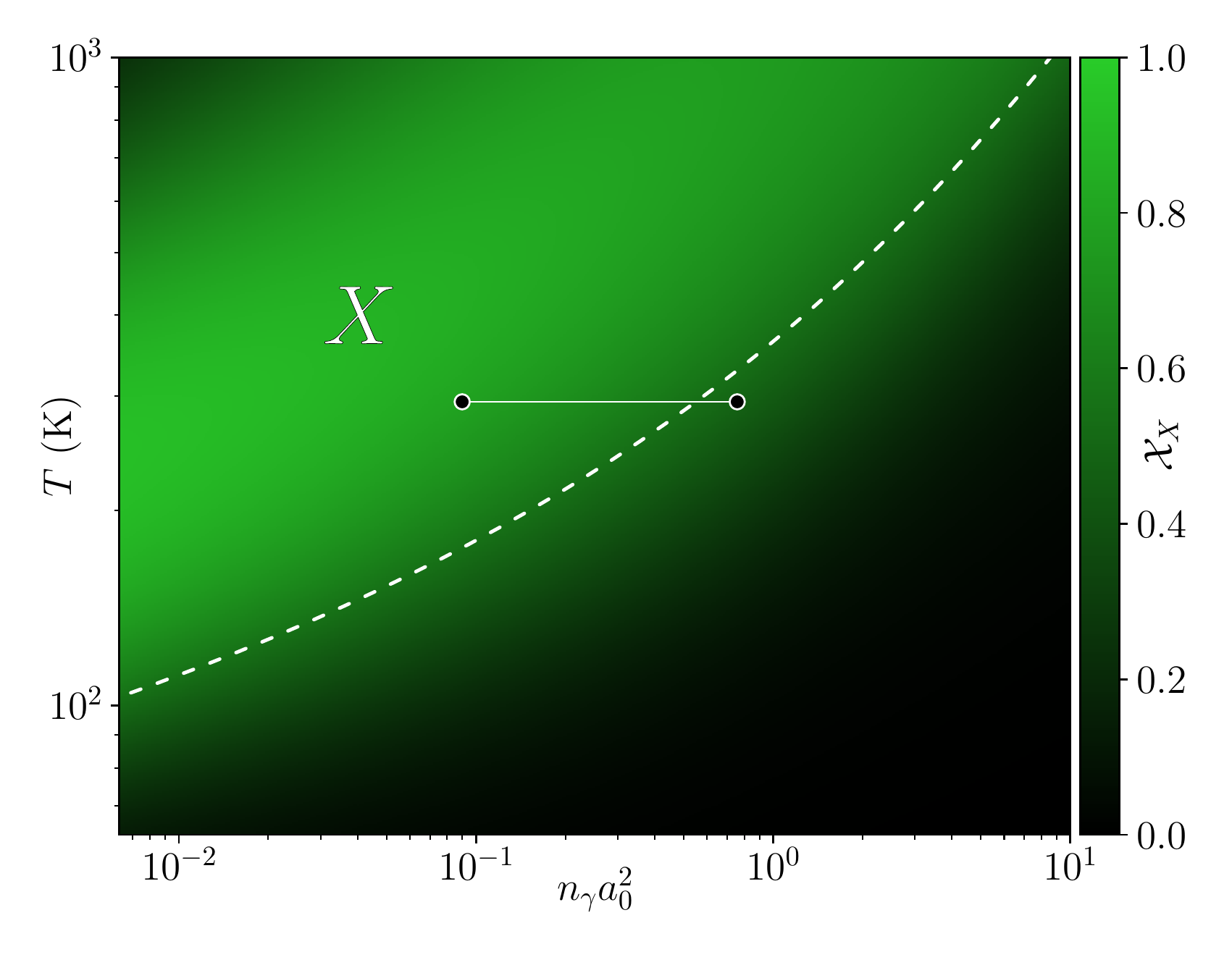}
        \caption{
            Fraction of excitons $\mathcal{X}_X$ as a function of temperature $T$ and photoexciation density scaled by the exciton Bohr area $n_\gamma a_0^2$.
            The dashed line is computed using Eqs. (\ref{eq:n_biexc_regime}) and (\ref{eq:n_biexc_regime_delta}).
            The maximum number of excitons per Bohr area is $n_X a_0^2 \simeq 3.30$, with $a_0^2 = 4.68$\ nm$^2$, obtained in the upper right corner.
        }
        \label{fig:diagram_n_X}
    \end{center}
\end{figure}

\begin{figure}[h!]
    \begin{center}
        \includegraphics[width=\linewidth]{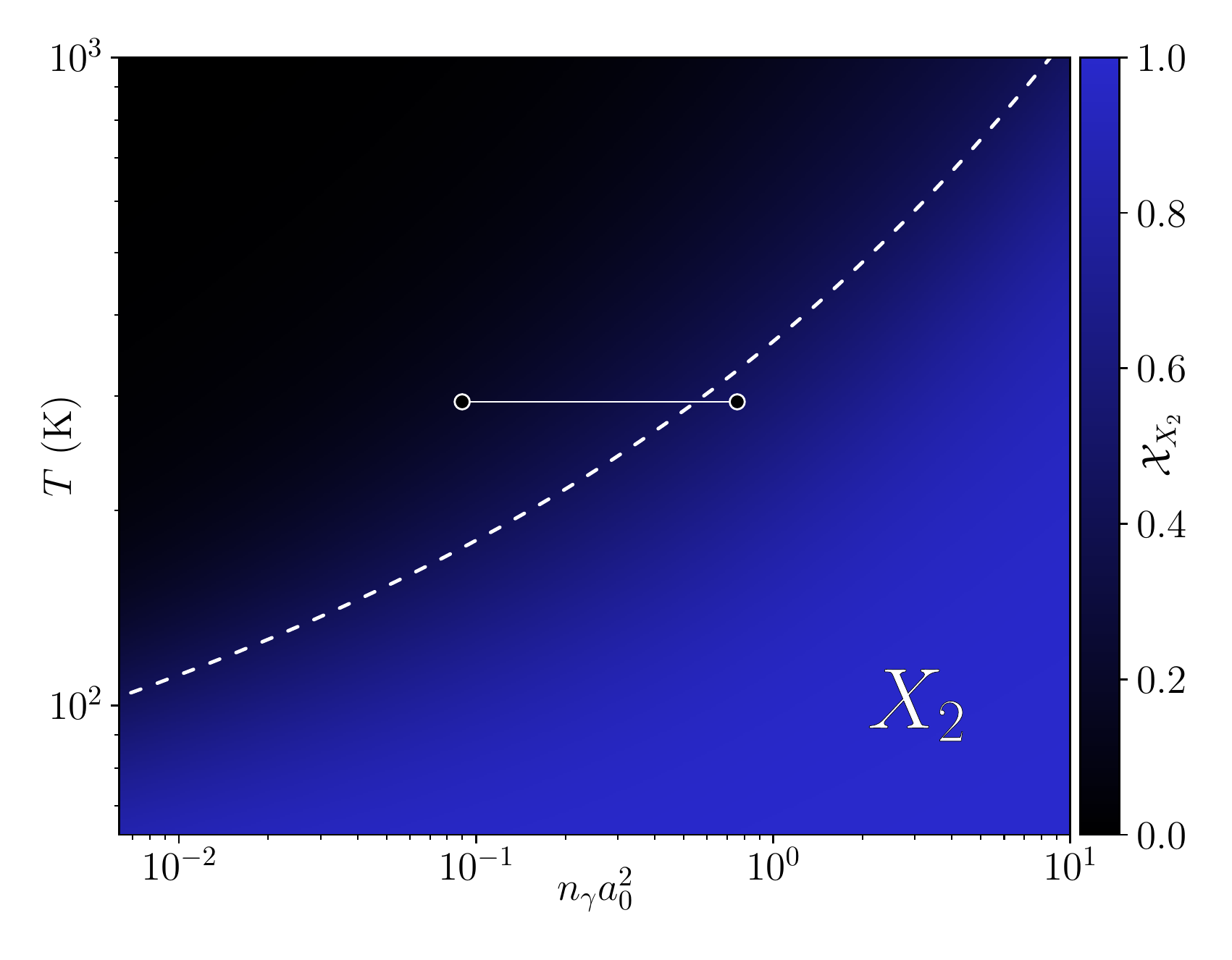}
        \caption{
            Fraction of biexcitons $\mathcal{X}_{X_2}$ as a function of temperature $T$ and photoexciation density scaled by the exciton Bohr area $n_\gamma a_0^2$.
            The dashed line is computed using Eq. (\ref{eq:n_biexc_regime}).
            The maximum number of biexcitons per Bohr area is $n_{X_2} a_0^2 \simeq 5.00$, with $a_0^2 = 4.68$\ nm$^2$, obtained in the lower right corner.
        }
        \label{fig:diagram_n_X_2}
    \end{center}
\end{figure}

Keeping that in mind, in Fig. \ref{fig:diagram} there are three clearly defined regions where each one of the three species dominates: $q$, $X$, and $X_2$; as well as two crossovers connecting them: $q \leftrightarrow X$ and $X \leftrightarrow X_2$.
Notice in particular that as the photoexcitation density increases, or the temperature lowers, a biexciton-dominated regime is always reached, which is a direct consequence of the stability of excitons, and consequently biexcitons, in CdSe nanoplatelets.
In order to study the behavior of each species in detail, Figs. \ref{fig:diagram_n_q}, \ref{fig:diagram_n_X}, and \ref{fig:diagram_n_X_2} respectively show the fraction of free charges $\mathcal{X}_q$, that of excitons $\mathcal{X}_X$, and that of biexcitons $\mathcal{X}_{X_2}$.

Figure \ref{fig:diagram_n_q} shows that the contribution from free charges to the overall picture is effectively negligible for most of the densities and temperatures considered.
Free charges are the dominant species only in the upper left corner, however notice that the highest density $n_q$ is found in the upper right corner, i.e., at high photoexcitation density and high temperature.
Furthermore, the region of the phase diagram that we are most interested in is the region around room temperature or below, in which biexcitons are mostly formed.
Since any screening, which would lower the energy of the exciton $E_X$ or biexciton $E_{X_2}$, is relatively constant at these temperatures due to the saturation of free charges, it may be effectively taken into account by using a slightly reduced value of $E_X$ and $E_{X_2}$, obtained either from experiments or theoretical calculations \cite{garciaflorez2019}.
Therefore such a change in the energy would in principle slightly shift the regions towards the left and up, since excitons and biexcitons would be easier to create.

Figure \ref{fig:diagram_n_X} shows noticeably different results at high photoexcitation density when compared with the biexciton-less model.
Due to the presence of biexcitons, the density of excitons also saturates once the biexciton-dominated regime is reached, analogously to the behavior of free charges in the $q \leftrightarrow X$ crossover.
This is a consequence of the energy level of the biexciton state being more negative than that of two unbound excitons, and thus at high photoexcitation density or low temperature, free charges mostly bind together into biexcitons.
In terms of the equation of state, Eq. (\ref{eq:id_gas_n_exc}) shows that $n_X$ saturates as the chemical potential approaches $\mu_X \rightarrow E_X + E_{X_2} / 2$, while $n_{X_2}$ diverges in that same limit, as given by Eq. (\ref{eq:id_gas_n_biexc}).
In this limit the saturated density of excitons $n_X^\infty$ is defined as

\begin{equation} \label{eq:n_exc_saturation}
    n_X^\infty \equiv -g_s^2 \frac{m_X k_B T}{2\pi \hbar^2} \ln\left(1 - e^{E_{X_2} / 2 k_B T}\right)
    ~~.
\end{equation}

\noindent Furthermore, we estimate the photoexcitation density at which the crossover $X \leftrightarrow X_2$ occurs, denoted by $n_\gamma^{c}$, by computing $n_{X_2} = n_X$.
This equality results in

\begin{align} \label{eq:n_biexc_regime}
    \begin{split}
        n_\gamma^{c} =& -g_s^2 \frac{m_X k_B T}{2\pi \hbar^2} \ln\left(1 - e^{\left(E_{X_2} - \delta\right) / k_B T}\right) \\
                        & -\frac{m_{X_2} k_B T}{\pi \hbar^2} \ln\left(1 - e^{-\delta / k_B T}\right) ~~, \\
    \end{split}
\end{align}

\noindent with $\delta$ given by

\begin{equation} \label{eq:n_biexc_regime_delta}
    \delta = 2 k_B T \ln\left[\cosh\left(\frac{E_{X_2}}{2 k_B T}\right)\right]
\end{equation}

\noindent for a certain temperature $T$.
Notice that the contribution from the density of free charges $n_q$ to the photoexcitation density is neglected here, since in the region $n_X = n_{X_2}$ it is valid to use that $n_q \ll n_X + 2 n_{X_2}$ as Fig. \ref{fig:diagram_n_q} shows.
Figures \ref{fig:diagram}, \ref{fig:diagram_n_X}, and \ref{fig:diagram_n_X_2} show a dashed line marking the crossover $X \leftrightarrow X_2$, from the exciton-dominated regime to the biexciton-dominated regime computed using Eqs. (\ref{eq:n_biexc_regime}) and (\ref{eq:n_biexc_regime_delta}).

Lastly, Fig. \ref{fig:diagram_n_X_2} clearly shows that biexcitons always dominate at low temperature or high photoexcitation density.
Notice that Fig. \ref{fig:diagram_n_X_2} shows the number of photoexcitations per Bohr area $n_\gamma a_0^2$ up to a value of 10, that is, the number of electrons per Bohr area, consequently the maximum amount of biexcitons per Bohr area is only but half of that.

In order to understand the behavior of excitons and biexcitons at high densities from a different angle, we introduce the thermal de Broglie wavelength for a species $\mathcal{Q}$, denoted by $\lambda_{\mathcal{Q}}$ and defined as

\begin{equation} \label{eq:lambda_th}
    \lambda_{\mathcal{Q}} \equiv \sqrt{\frac{2 \pi \hbar^2}{m_{\mathcal{Q}} k_B T}}
    ~~,
\end{equation}

\noindent where $m_{\mathcal{Q}}$ is either the mass of the exciton or that of the biexciton, with $\hbar$ the reduced Planck constant.
Recall that physically $\lambda_{\mathcal{Q}}$ is the de Broglie particle size, which means that quantum effects are important when the inter-particle distance is of the order of $\lambda_{\mathcal{Q}}$.
In other words, when the number of either excitons or biexctions per thermal wavelength squared, that is, $n_{X} \lambda_{X}^2$ or $n_{X_2} \lambda_{X_2}^2$, becomes of the order of the number of spin states with the same energy for that particular species, they are considered to be ``quantum degenerate'' and hence described by the Bose-Einstein distribution.
Thus, in order to account for the spin degrees of freedom we define the degeneracy parameter $\eta_{\mathcal{Q}}$ as

\begin{equation} \label{eq:deg_parameter_def}
    \eta_{X} \equiv \frac{n_X \lambda_{X}^2}{g_s^2}
    ~~ , \text{ and } ~~
    \eta_{X_2} \equiv n_{X_2} \lambda_{X_2}^2
    ~~ ,
\end{equation}

\noindent where $g_s^2 = 4$ for excitons.
As $\eta_{\mathcal{Q}}$ approaches one, a correct description for the behavior of species $\mathcal{Q}$ requires the use of quantum statistics, i.e., Fermi-Dirac for fermions and Bose-Einstein for bosons.

\begin{figure}[h!]
    \begin{center}
        \includegraphics[width=\linewidth]{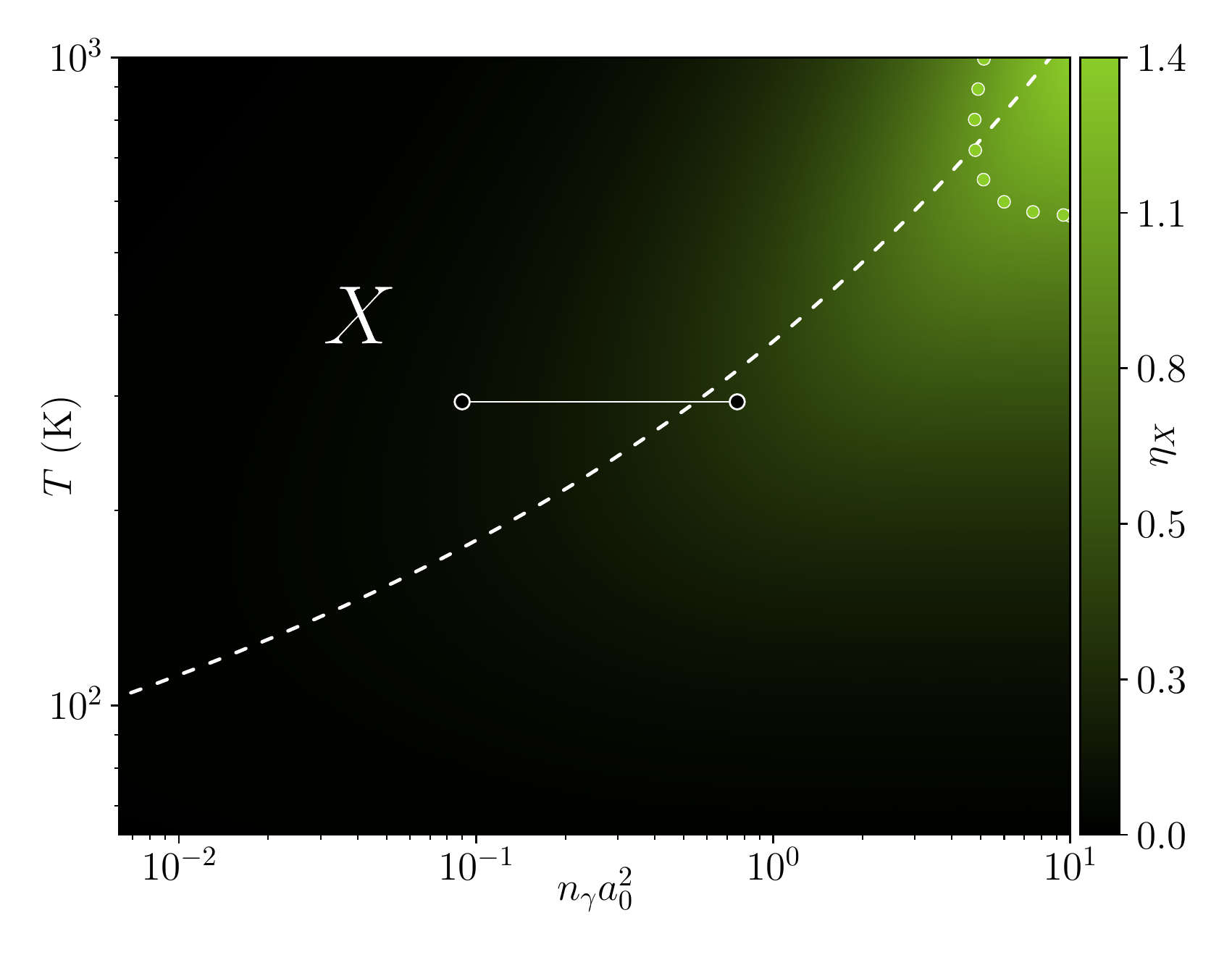}
        \caption{
            Degeneracy parameter for excitons $\eta_X \equiv n_X \lambda_{X}^2 / g_s^2$ as a function of temperature $T$ and photoexciation density scaled by the exciton Bohr area $n_\gamma a_0^2$.
            The circles (\excdeg) dotted line represents the photoexcitation density for which $\eta_X = 1$, computed using Eq. (\ref{eq:n_exc_deg}).
            The minimum degeneracy temperature $T^*$, computed using Eq. (\ref{eq:T_exc_deg}), is $T^* \simeq 569$\ K.
            The maximum value of the degeneracy parameter is $\eta_{X} \simeq 1.36$.
        }
        \label{fig:diagram_n_X_deg}
    \end{center}
\end{figure}

Let us analyze the degeneracy parameter of excitons first.
Figure \ref{fig:diagram_n_X_deg} shows $\eta_X$ as a function of temperature $T$ and photoexcitations per Bohr area $n_\gamma a_0^2$.
At first glance it is noticeable that, compared with Fig. \ref{fig:diagram_n_X}, the regime in which excitons become degenerate does not align with the one in which they are dominant.
Naturally, this comes as a consequence of the actual density of excitons being high enough only in a small region inside the biexciton-dominated regime for the temperatures considered.

An estimate of the photoexcitation density at which excitons behave as quantum particles, denoted by $n_\gamma^{\excdeg}$, is obtained by finding the density of excitons that satisfy the equality $\eta_X = 1$.
With this value for the density of excitons, we obtain that the corresponding photoexcitation density is

\begin{align} \label{eq:n_exc_deg}
    \begin{split}
        n_\gamma^{\excdeg} &= g_s^2 \frac{m_X k_B T}{2\pi\hbar^2} \\
                                &- \frac{m_{X_2} k_B T}{2 \pi \hbar^2} \ln \left(1 - \left( 1 - \frac{1}{e} \right)^2 e^{-E_{X_2} / k_B T} \right) ~~ ,
    \end{split}
\end{align}

\noindent for a certain temperature $T$.
Figure \ref{fig:diagram_n_X_deg} shows $n^{\excdeg}_\gamma$ as a circle (\excdeg) dotted line.
Interestingly, there is a minimum temperature $T^*$ below which excitons do not become degenerate, as their density does not reach the threshold $\eta_X = 1$ since at lower temperatures than $T^*$ charges mostly bind into biexcitons.
Solving for the temperature at which the argument of the logarithm in Eq. (\ref{eq:n_exc_deg}) becomes zero, we obtain that valid temperatures satisfy

\begin{equation} \label{eq:T_exc_deg}
    T > \frac{E_{X_2}}{2k_B} \frac{1}{\ln\left(1 - \frac{1}{e}\right)} \equiv T^*
    ~~ .
\end{equation}

\noindent Notice that it only depends on the energy level of the biexciton state $E_{X_2}$, and consequently $T^*$ approaches zero in the limit $E_{X_2} \rightarrow 0$, i.e., when there is no biexciton state.

\begin{figure}[h!]
    \begin{center}
        \includegraphics[width=\linewidth]{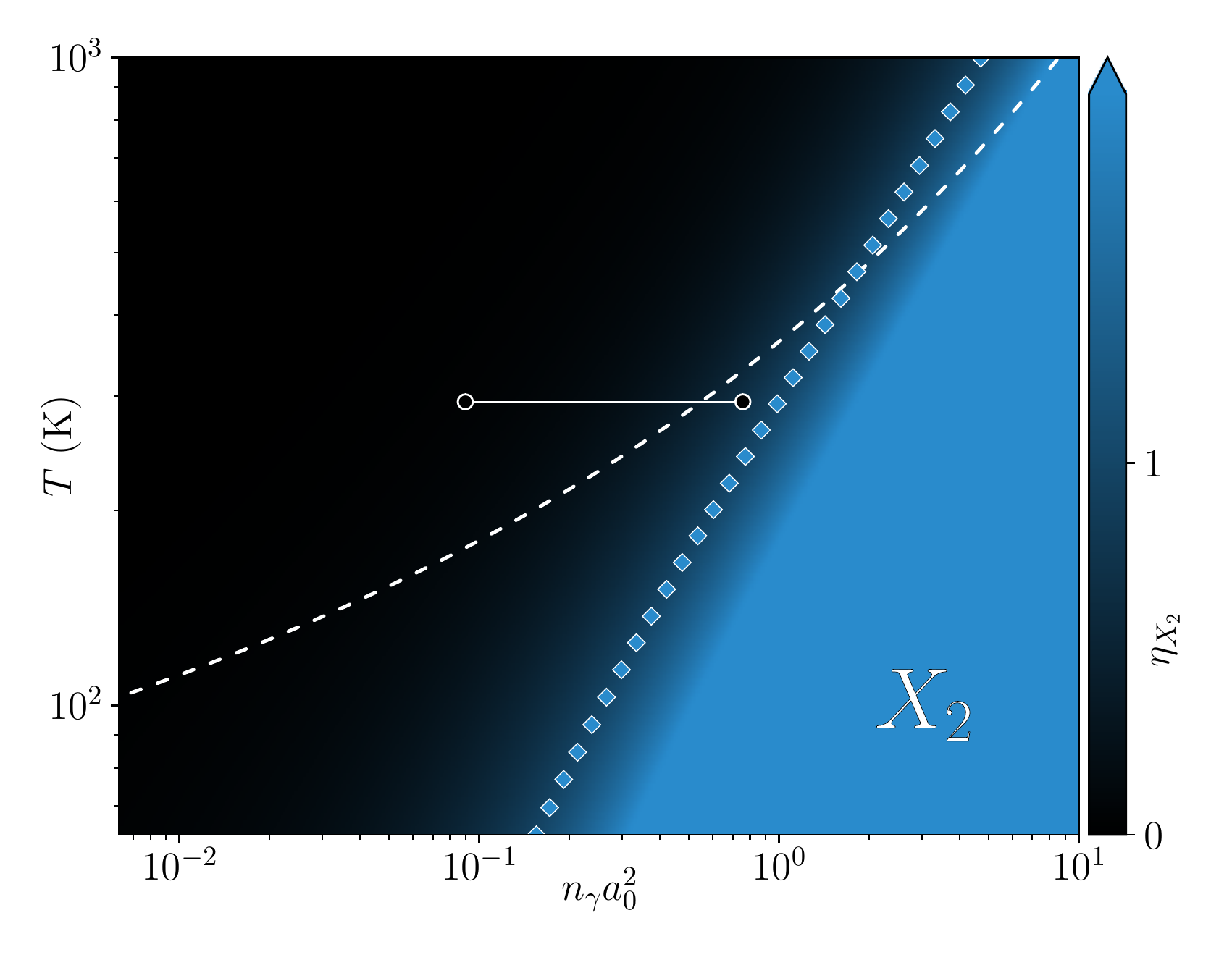}
        \caption{
            Degeneracy parameter for biexcitons $\eta_{X_2} \equiv n_{X_2} \lambda_{X_2}^2$ as a function of temperature $T$ and photoexciation density scaled by the exciton Bohr area $n_\gamma a_0^2$.
            The squares (\biexcdeg) dotted line represents the photoexcitation density for which $\eta_{X_2} = 1$, computed using Eq. (\ref{eq:n_biexc_deg}).
            The maximum value of the degeneracy parameter is $\eta_{X_2} \simeq 65.3$, however points for which $\eta_{X_2} \geq 2$ are represented with the same color.
        }
        \label{fig:diagram_n_X_2_deg}
    \end{center}
\end{figure}

Regarding the degeneracy parameter of biexcitons, shown in Fig. \ref{fig:diagram_n_X_2_deg}, notice that the region in which they are degenerate is clearly very different from that of the excitons.
Naturally, since the energy of the biexciton state is more negative than that of two unbound excitons, at high photoexcitation density or low temperature there is always a degenerate regime.
As before, an analytical expression for the photoexcitation density above which the biexcitons behave as quantum particles, denoted by $n_\gamma^{\biexcdeg}$, is obtained by solving $\eta_{X_2} = 1$.
Analogously to the calculation for excitons, the corresponding result is

\begin{align} \label{eq:n_biexc_deg}
    \begin{split}
        n_\gamma^{\biexcdeg} = &-g_s^2 \frac{m_X k_B T}{2\pi\hbar^2} \ln\left(1 - \sqrt{1 - \frac{1}{e}} ~ e^{E_{X_2} / 2 k_B T}\right) \\
                                    & + \frac{m_{X_2} k_B T}{2 \pi \hbar^2} ~~ ,
    \end{split}
\end{align}

\noindent for a certain temperature $T$.
Notice that due to the argument of the logarithm, unlike for excitons, there is no minimum temperature below which biexcitons are not degenerate.
Physically, this different behavior is explained by the fact that in our model the ground state of the system corresponds to the free charges being completely bound into biexcitons.
Figure \ref{fig:diagram_n_X_2_deg} shows $n_\gamma^{\biexcdeg}$, computed using Eq. (\ref{eq:n_biexc_deg}), as a square (\biexcdeg) dotted line.
Notice that values $\eta_{X_2} \geq 2$ are represented with the same color, since biexcitons become heavily degenerate, that is, the maximum degeneracy obtained in Fig. \ref{fig:diagram_n_X_2_deg} is much larger than one.

\section{\label{sec:discussion}Conclusion and Outlook}

Summarizing, we have presented a thermodynamical model involving four ideal gases that results in a phase diagram describing the physics of free charges (electrons and holes), excitons, and biexcitons in CdSe nanoplatelets.
Because of the ideal gas description various interaction effects are implicitly neglected thus at high enough photoexcitation densities we expect this model to break down, and this we want to explore in more detail in the future.
Clearly the formation of excitons and biexcitons due to the attractive electron-hole interaction is considered.
However, self-energy effects on the electrons, holes, excitons and biexcitons, i.e., energy shifts and finite lifetime effects caused both by electron-hole attraction and electron-electron and hole-hole repulsion, are neglected.
In the specific case of interactions between free charges and excitons, leading in particular to screening effects, experimental results on CdSe nanoplatelets show that these may be neglected in the model by setting the energies $E_X$ and $E_{X_2}$ to constant values, at least for the range of density and temperature of interest \cite{garciaflorez2019}.
Only at high enough temperatures, i.e., higher than $T \simeq 10^3$\ K, the system reaches a free-charges-dominated regime and thus interactions that may form an exciton would be heavily screened.\
In conclusion, we have provided in this paper a comprehensive study of a thermodynamical model suitable for experimentally testable predictions, and that moreover may possibly also be used for guiding future endeavors both in experiments and theory involving biexcitons in two-dimensional materials other than CdSe nanoplatelets if the exciton binding energy is sufficiently large.
Of particular interest are for instance also PbS nanoplatelets due to the shared features with CdSe nanoplatelets, such as indeed a large exciton binding energy ($400$ to $500$\ meV), as well as similar synthesis procedures and optical properties \cite{lauth2019,li2016,bouet2014,sonntag2019,khan2017}.

\section*{Acknowledgments}

This work is part of the research programme TOP-ECHO with project number 715.016.002, and is also supported by the D-ITP consortium.
Both are programmes of the Netherlands Organisation for Scientific Research (NWO) that is funded by the Dutch Ministry of Education, Culture and Science (OCW).

%\newpage

%\appendix

%\nocite{*}
\bibliography{reference}% Produces the bibliography via BibTeX.

\end{document}